\documentclass[%
 aip,
 rsi,
 amsmath,amssymb,
reprint,%
]{revtex4-1}

\usepackage[dvipdfmx]{graphicx}
\usepackage{dcolumn}
\usepackage{bm}
\usepackage{amsmath, amssymb}

\usepackage[utf8]{inputenc}
\usepackage[T1]{fontenc}
\usepackage{mathptmx}

\begin{document}

\preprint{AIP/123-QED}

\title[Current-feedback-stabilized laser system for quantum simulation experiments using Yb clock transition at 578~nm]{Current-feedback-stabilized laser system for quantum simulation experiments using Yb clock transition at 578~nm}

\author{Y. Takata}
  \email{takata@yagura.scphys.kyoto-u.ac.jp}
  \affiliation{Department of Physics, Graduate School of Science, Kyoto University, 606-8502, Japan
}
\author{S. Nakajima}
  \affiliation{The Hakubi Center for Advanced Research, Kyoto University, Kyoto 606-8501, Japan
}
\author{J. Kobayashi}
  \affiliation{PRESTO, Japan Science and Technology Agency, Kyoto 606-8502, Japan
}
\author{K. Ono}
\author{Y. Amano}
\author{Y. Takahashi}
  \affiliation{Department of Physics, Graduate School of Science, Kyoto University, 606-8502, Japan
}

\date{\today}


\begin{abstract}
We developed a laser system for the spectroscopy of the clock transition in ytterbium (Yb) atoms at 578~nm based on an interference-filter stabilized external-cavity diode laser (IFDL) emitting at 1156~nm.
Owing to the improved frequency-to-current response of the laser-diode chip and the less sensitivity of the IFDL to mechanical perturbations, 
we succeeded in stabilizing the frequency to a high-finesse ultra-low-expansion glass cavity with a simple current feedback system.
Using this laser system, we performed high-resolution clock spectroscopy of Yb and found that the linewidth of the stabilized laser was less than 320~Hz.
\end{abstract}

\maketitle


\section{INTRODUCTION}
In recent years, significant attention has been given to the $^1S_0$-$^3P_0$ intercombination transitions or ''clock transitions'' of alkaline-earth like atoms owing to their various applications, 
including optical lattice clocks\cite{Katori2003}$^,$\cite{Mcgrew2018}, quantum simulators\cite{Gorshkov2010}$^,$\cite{Ono2019}, and quantum information processing\cite{Daley2011}.
In order to use such clock transitions, 
the development of an ultranarrow laser is crucially important due to the narrow linewidth of the $^1S_0$-$^3P_0$ transition.

For the clock transition of ytterbium (Yb) atoms at 578~nm, 
dye-lasers\cite{Hoyt2005} and diode lasers with sum-frequency generations\cite{Inaba2013} are used to excite this transition.
More recently, diode lasers with second-harmonic generations (SHG) have become widely used\cite{Nevsky2008}$^,$\cite{Lee2011}$^,$\cite{Cappellini2015}$^,$\cite{Kobayashi2016} owing to their simplicity over other schemes.
A quantum-dot laser diode at 1156~nm used in previous reports, however, has unfavorable characteristics, namely, the response of the laser frequency against the laser-diode current modulation at high frequency is slow\cite{Cappellini2015} 
and an electro-optic modulator (EOM) or special electrical filter must be employed for the feedback of the fast error signal.

In this study, we report a versatile laser system that enables us to excite the Yb $^1S_0$-$^3P_0$ clock transition at 578~nm with a simple current feedback stabilization system.
As a key component of the system, we constructed an interference-filter stabilized external-cavity diode laser\cite{Baillard2006} (IFDL) emitting at 1156~nm, 
the laser frequency of which is less sensitive to mechanical perturbations. 
A 578~nm light is generated through SHG using a waveguide of periodically poled lithium niobate (PPLN).
In this study, we adopted an improved quantum-dot LD whose internal waveguide stripe differs from those used in previous reports.
We measured the frequency-to-current response of the gain chip and found that the chip has a fast response comparable to a normal laser diode.
This enabled us to stabilize the laser frequency to an ultra-low expansion (ULE) glass cavity with a finesse of approximately 60,000 using typical PID current feedback without the application of a special optical or electrical component. 
Using this stabilized laser, we demonstrated the spectroscopy of the $^1S_0$-$^3P_0$ transition in Yb and found that the laser linewidth was less than 320~Hz at 578~nm.


\section{FILTER-STABILIZED DIODE LASER}
\begin{figure}[tbp]
\includegraphics[width=8.0cm, clip]{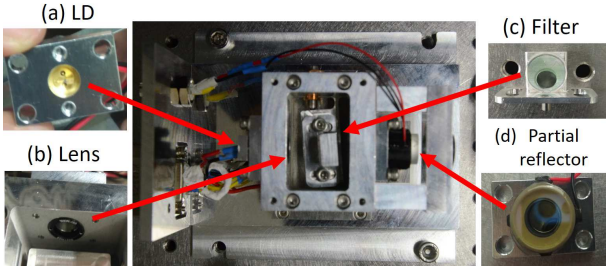}
\caption{ Photograph of the constructed IFDL. The set of mechanical parts is made of aluminum and sold by Samplus Trading Inc.
The 1156~nm light generated by the LD (a) is collimated by the first collimation lens (b). The collimated beam then passes through the interference filter (c) for the wavelength selection and is focused on the retroreflective mirror (d) by the second collimation lens mounted in the same way as the first lens (b).} 
\label{fig:IFDL}
\end{figure}

The design of our IFDL is shown in Fig.\ref{fig:IFDL}.
The output of a laser diode (LD) is retroreflected by a partial reflective mirror in a cateye configuration, which forms a laser cavity, and the wavelength of the IFDL is selected by tuning the angle of the interference filter inside the cavity.
The retroreflector placed in a cateye configuration makes the laser frequency less sensitive to mechanical perturbations.
This characteristic is beneficial, particularly for the excitation of narrow transitions such as clock transitions.
We adopted a quantum-dot LD (Innolume GC-1180-100-TO-200-B) with a single AR-coating facet.
Different from a previous product (Innolume GC-1156-TO-200), this LD has a curved waveguide that suppresses lasing by itself.
The injection current of the LD is controlled using a commercial current controller (Vescent Photonics D2-105-500) and a servo system (Vescent Photonics D2-125), the power of which is supplied by a low-noise power supply (Vescent Photonics D2-005).
The output beam of the LD is collimated by an aspherical lens.
The collimated beam passes through an interference filter and is focused on a partially reflective mirror (output coupler) by another aspherical lens in a cateye configuration.
The output coupler has a reflectance of 25~\% and is mounted on a ring piezo actuator 
(Piezomechanik HPSt 500/15-8/5).
The cavity length is about 5~cm, which can be fine-tuned using the piezo actuator.
The laser frequency is selected by an interference filter mounted on an stage with a variable angle in the cavity.
The maximum transmittance of the filter is 72~\% at 1157.6~nm and the full width at half maximum (FWHM) of the transmittance spectrum is 0.44~nm.
The angle of the filter is fixed around $4^\circ$ for lasing at 1156~nm.
The transmitted light of the output coupler is collimated again by an aspherical lens.
The temperature of all components is stabilized using a single Pertier device.

\begin{figure}[tbp]
\includegraphics[width=8.2cm,clip]{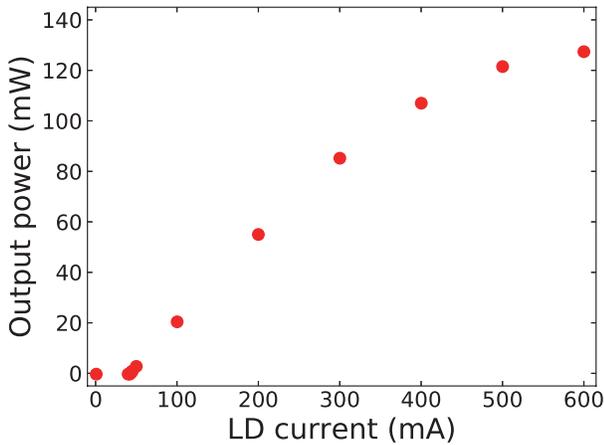}
\caption{Output power of the IFDL as a function of the laser-diode current.
}
\label{fig:P-I}
\end{figure}

We can obtain an output power of 127~mW with an injection current of 600~mA 
(Fig.\ref{fig:P-I}).
The output power of the IFDL is slightly weaker than the value on the data sheet taken with Littrow ECDL with 10~\% optical feedback. 
The frequency stabilization becomes unstable at above a current of 400~mA. 
Considering the saturating behavior at over 400~mA, such instability can be attributed to the over injection to the laser diode. 

\section{FREQUENCY-TO-CURRENT RESPONSE}
In previous reports using a commercial laser diode at 1156~nm\cite{Lee2011}$^,$\cite{Cappellini2015}$^,$\cite{Kobayashi2016}, 
GC-1156-TO-200 (Innolume) was adopted as a gain chip.
Strangely, the chip has a slow frequency-to-current response compared to typical laser diodes. 
According to Ref.\onlinecite{Cappellini2015}, the phase of the frequency response flips from 0$^\circ$ to 180$^\circ$ at a current modulation frequency of between 10 and 100~kHz current modulation frequency.
This characteristic severely limits the feedback bandwidth.
To overcome this problem, another actuator such as an EOM or special electrical filter for the feedback of a fast error signal was employed.

Thanks to the recent development of a new manufacturing process for laser waveguides, 
other types of laser diodes have become commercially available for the 1156-nm wavelength.
In this study, we explore the possibility of using a GC-1180-100-TO-200-B (Innolume), which has a curved waveguide to suppress lasing by itself.

\begin{figure}[tbp]
\includegraphics[width=8.0cm,clip]{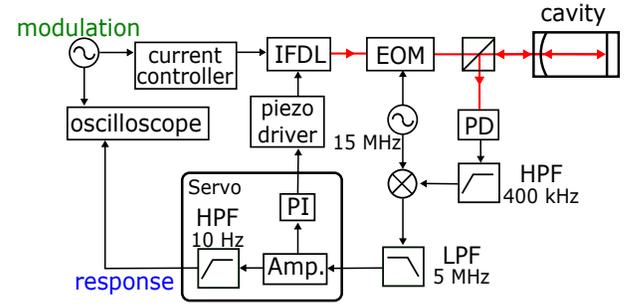}
\caption{Experimental setup used for frequency-to-current response measurement.
The error signal generated via Pound-Drever-Hall scheme is fed back to the piezo actuator in the IFDL. 
The modulation applied to the laser-diode current causes an oscillation of the error signal, from which we evaluate the current response.}
\label{fig:measure}
\end{figure}

\begin{figure}[tbp]
\includegraphics[width=7.2cm, clip]{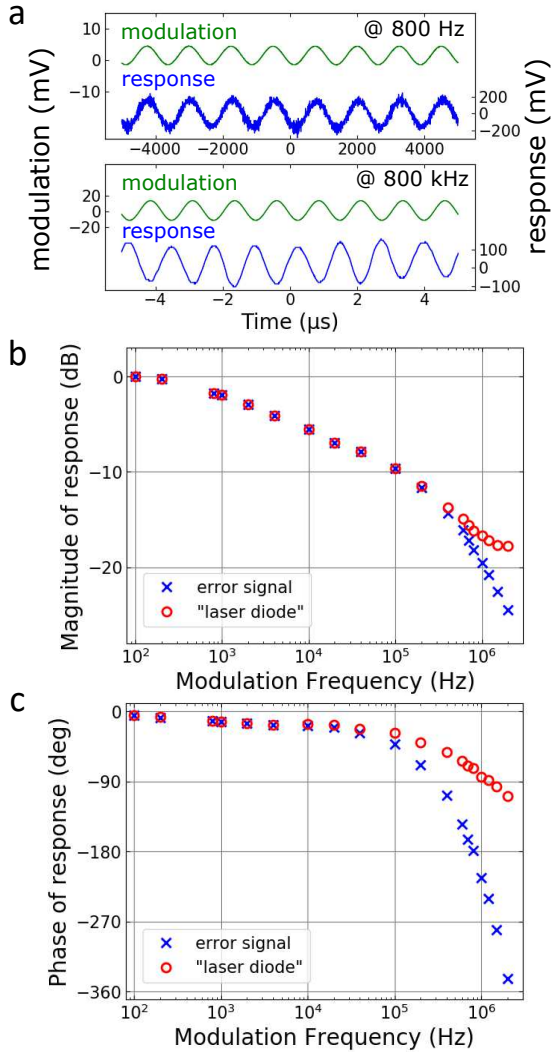}
\caption{Measurement of the frequency-to-current response. 
(a) Signals measured using an oscilloscope.
The acquisition of the signals was triggered by a modulation signal, and the signals were averaged fro over 128 shots.
The response is in phase with the modulation at 800~Hz (upper image).
By contrast, the response is delayed by 180$^\circ$ in phase with the modulation at 800~kHz (lower image).
(b) Magnitude of the response. The data of the error signal (blue crosses) are the responses measured in the setup shown in Fig.\ref{fig:measure}. 
The data of the ``laser diode'' (red circles), which should reflect the response of the laser diode, were obtained by subtracting the contributions other than the LD from the obtained error signals (blue crosses). 
The values were normalized to the response amplitude at 100-Hz modulation. 
(c) The phase angle of the response relative to the modulation. The phase of the LD response crosses the 90$^\circ$ line at a modulation frequency.of approximately  1.2~MHz. }
\label{fig:response}
\end{figure}

To evaluate the feedback properties of this new product, we measured the frequency-to-current response.
The measurement setup is shown in Fig.\ref{fig:measure}.
For this measurement, we constructed a low-finesse cavity whose finesse and FWHM at 1156~nm are 
$\sim$600 and $\sim$2~MHz, respectively.
The frequency response was evaluated using an error signal produced by the Pound-Drever-Hall (PDH) scheme\cite{PDH1983}.
To generate frequency sidebands for the PDH scheme, an EOM driven at a 15-MHz radio-frequency (RF) from a local oscillator was used.
The RF signal from the photodiode passing through a high-pass filter with a cutoff frequency of 400~kHz 
was mixed into the local oscillator signal and filtered by a low-pass filter with a cutoff frequency of 5~MHz to generate an error signal.
Note that the laser frequency was kept near the resonance of the cavity through the weak and slow feedback of the error signal to the piezo actuator.
This piezo feedback is done below the 100~Hz frequency range where we expect the normal response of the laser frequency versus LD current to be located.
To determine the response of the LD, 
we added a small LD current modulation of less than 40~$\mu$A to the 300~mA direct current.
By comparing the modulation signal and modulated error signal, shown in Fig.\ref{fig:response}(a),
we can evaluate the response of the error signal against the LD current modulation.

The amplitude and phase of the obtained responses are shown in Figs.\ref{fig:response}(b) and \ref{fig:response}(c), respectively.
The observed response comes from not only the LD response itself but also that of other optical or electrical components.
Thus, to solely obtain the LD response, we subtract the contributions from other parts.
One contribution comes from the response of the PDH system.
The response of the error signal in the PDH scheme is similar to that of a low-pass filter with a cutoff frequency of 1.05~MHz, 
corresponding to the half-width at half-maximum\cite{Rakhmanov2002} (HWHM) of the used cavity measured from the transmission signal.
Another contribution is the electric and optical propagation delay.
By summing the delays of the electronic low-pass filter, laser servo, coaxial cables, optical fiber, and optical path in the space, 
we estimate the propagation delay to be 240~ns.
Subtracting these contributions, 
we can conclude that the phase flip of the LD occurs around 1.2-MHz modulation, which is one order of magnitude larger than that of previous reports.
Note that we do not consider the laser linewidth or propagation delays from other components in this analysis.
Furthermore, the weak and slow feedback to the piezo actuator is not included in the analysis.
We measured the response at a modulation frequency of 100~Hz and found that the amplitude of the response is suppressed by only 3~\%.
Because the resonance of the piezo actuator cannot be seen in the response, we can conclude that the effect of the feedback to the measurement is negligibly small. 

In order to check the consistency of the response measurement, 
we locked the laser frequency to the cavity resonance by the feedback to the laser-diode current in the setup. 
At a large feedback gain, a peak was observed at approximately 800~kHz in the error signal spectrum.
This is consistent with the response measurement where the phase of the measured response crosses the 180$^\circ$ line at 800~kHz.


\section{SECOND HARMONIC GENERATION}

To generate 578~nm light from the 1156~nm light, we use a waveguide PPLN crystal manufactured by NTT Electronics.
This is the same product as that used in Ref.\onlinecite{Kobayashi2016}.
The 1156~nm beam passes through two optical isolators with an isolation of 77~dB and is then coupled to the PPLN through a polarization-maintaining fiber.
Under typical conditions, the input power of the 1156~nm laser beam before the fiber coupler of the PPLN module is 80~mW and a total of 26~mW of the 1156~nm laser power couples to the PPLN. 
We can obtain an 18-mW output power at 578~nm under the optimum temperature.


\section{LOCKING TO THE ULE CAVITY}

\begin{figure}[tbp]
\includegraphics[width=8.0cm,clip]{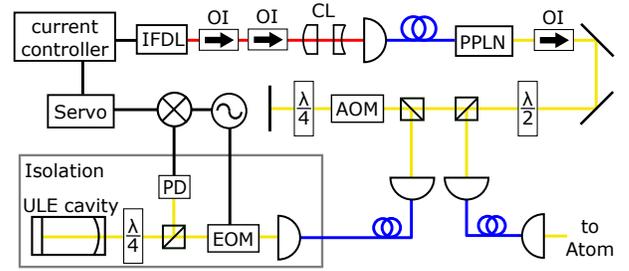}
\caption{Schematic drawing of clock laser system for spectroscopy.
After passing through two optical isolators (OI), the beam circularized by two cylindrical lenses (CL) was coupled to the fiber connected to the waveguide PPLN. 
A portion of the 578-nm laser is brought to a high-finesse ULE cavity for the frequency stabilization.
In order to compensate the frequency difference between the atomic transition and the cavity resonance, 
the laser frequency is shifted using a double-pass acousto-optic modulator (AOM).
The laser is stabilized by a PDH scheme through feedback to the LD current.}
\label{fig:overview}
\end{figure}

To obtain the short- and long-term stability of the laser frequency, 
we stabilized our laser to a high-finesse cavity made of ULE glass as a stable frequency reference.
The ULE cavity was placed in a vacuum chamber whose temperature was stabilized using a Pertier device attached to the outside of the chamber.
To isolate it from mechanical vibrations, the chamber was placed on a passive vibration-isolation table surrounded by an acoustic isolation box.
The finesse of the cavity was about 60,000 for the wavelength of 578~nm, meaning that the HWHM of the cavity resonance spectrum was about 25~kHz.

A schematic drawing of the locking system for the clock laser is shown in Fig.\ref{fig:overview}.
A portion of the 578-nm laser generated at the waveguide PPLN is used for the frequency stabilization.
In order to compensate the frequency difference between the atomic transition and the cavity resonance, 
the laser frequency is shifted using a double-pass acousto-optic modulator (AOM).
The error signal is generated by a PDH scheme and is fed back to the LD current through the same current controller and servo as used in the frequency-to-current response measurement.
We succeeded in stabilizing the laser to the ULE cavity through feedback to the LD current without additional stabilization schemes such as a pre-stabilizer\cite{Rohde2002}, AOM or EOM.
The remainder of the 578-nm laser power is brought to the atomic experiment through a 25-m fiber with fiber noise cancellation\cite{FNC1994}.

\begin{figure}[tbp]
\includegraphics[width=8.0cm, clip]{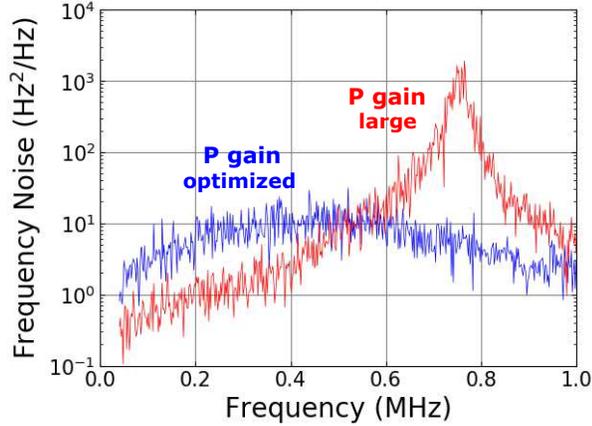}
\caption{Power spectrum of the error signal when the laser is stabilized to the high-finesse cavity. The blue and red signals show the spectrum when the PID parameters are optimum and the proportional gain is larger than the optimized point, respectively.
Since the oscillation around the 750~kHz appeared when the feedback gain is large, we estimate the feedback bandwidth is about 750~kHz.}
\label{fig:error}
\end{figure}

Figure \ref{fig:error} shows the power spectrum of the error signal. 
When the PID parameters of the servo are optimized, the spectrum has no large peaks. 
As we increase the proportional gain from the optimized parameters, 
the system starts to oscillate and a large peak around 750~kHz appears in the spectrum. 
Based on the above, we estimated the feedback bandwidth to be approximately 750~kHz. 
The reduction of the feedback bandwidth from the test system with the lower finesse was caused by the narrower width of the cavity resonance.
We achieved stable frequency stabilization without any relocking the system for the time duration typically longer than one day.

\section{Y\lowercase{b} SPECTROSCOPY}

\begin{figure}[tbp]
\includegraphics[width=7.5cm,clip]{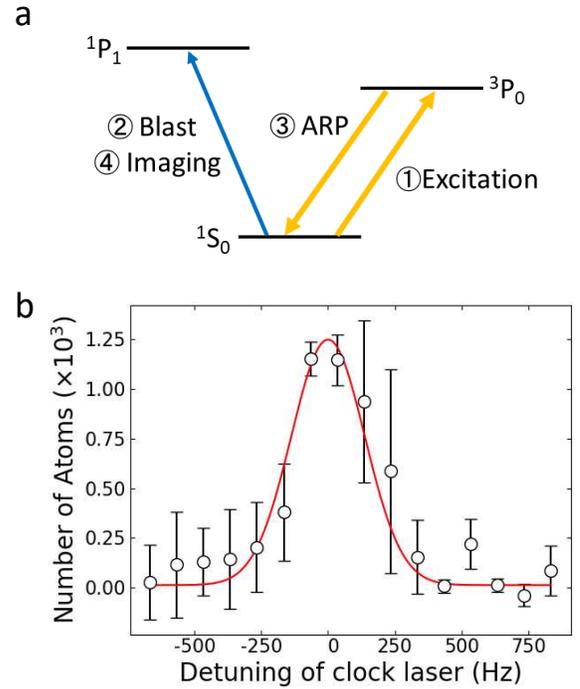}
\caption{Clock transition spectroscopy for $^{171}$Yb atoms in an optical lattice with the magic-wavelength. 
(a)Spectroscopy procedure after loading the atoms into the lattice. 
The 578-nm laser is irradiated to the atoms for 100~ms and a part of the atoms in the $^1S_0$ state are excited into the $^3P_0$ state.
The atoms remaining in the $^1S_0$ state are removed by the blasting light resonant to the $^1S_0$-$^1P_1$ transition.
After the removal, the atoms in the $^3P_0$ state are returned to the $^1S_0$ state by the 578-nm laser with the adiabatic rapid passage (ARP).
Then, the number of atoms in $^1S_0$ is measured via absorption imaging with the $^1S_0$-$^1P_1$ transition.  
(b) The obtained spectrum. A solid line indicates a Gaussian function generated by fitting to the data. Error bars are the standard deviations of the mean values of three measurements. From this measurement, the linewidth of the laser is deduced to be less than 320~Hz.
}
\label{fig:spectroscopy}
\end{figure}

In order to evaluate the linewidth of the 578~nm laser, we performed a spectroscopy of the clock transition with $^{171}$Yb in an optical lattice.
The detailed experimental setup is written in Ref.\onlinecite{Ono2019} and we describe it in short here.
We prepare the degenerate $^{171}$Yb atoms in a crossed-optical trap by a sympathetic evaporative cooling with $^{173}$Yb atoms.
After the removal of $^{173}$Yb, $^{171}$Yb atoms are loaded into the 3D optical lattice with the magic wavelength of about 759~nm, which gives the same trapping potential for the $^1S_0$ and $^3P_0$ states.
In this configuration, we can strongly suppress the many types of spectral broadening effects such as the Doppler effect, the differential light shift, and the collisional shift, as is well known in discussions on optical lattice clocks\cite{Katori2003}.
The depth of the lattice is 30 $E_R$, where $E_R$ is the recoil energy of the lattice laser, which is sufficiently deep to be within the Lamb-Dicke regime.
A magnetic field of 15~G is applied to spectroscopically separate the nuclear spin components.
The 578-nm laser with $\pi$-polarization is irradiated to the atoms for 100~ms and a part of atoms in the $^1S_0$ state are excited into the $^3P_0$ state.
The atoms remaining in the $^1S_0$ state are removed by the light resonant to the $^1S_0$-$^1P_1$ transition.
After the removal, the atoms in the $^3P_0$ state are returned to the $^1S_0$ state by the 578-nm laser with an adiabatic frequency sweep.
The number of atoms in the $^1S_0$ state is then measured via absorption imaging with the $^1S_0$-$^1P_1$ transition.

The spectrum obtained is shown in Fig.\ref{fig:spectroscopy}.
The Gaussian fitting to the data gives the FWHM of the spectrum of 324~Hz.
The intensity of the 578-nm laser is 0.3~mW/cm$^2$ and
its power broadening is estimated to be about 100~Hz. 
Taking this into consideration, the linewidth of the laser is deduced to be less than 320~Hz, which is narrow enough for most quantum simulation experiments\cite{Ono2019}$^,$\cite{Kato2016}$^,$\cite{Nakamura2019}.


\section{CONCLUSION}
We constructed an external-cavity diode laser system with an interference filter
for the clock transition of ytterbium atoms and obtained an output of 18~mW output at 578~nm.
We adopted a new type of LD chip for 1156~nm and showed that it has a fast frequency-to-current response unlike the LDs in the previous reports.
We succeeded in stabilizing the frequency of the laser to the high-finesse ULE cavity by feedback to the LD current without special electrical or optical elements.
The feedback bandwidth of the stabilization is about 750~kHz. 
Using the stabilized laser, we succeeded in high-resolution spectroscopy of $^{171}$Yb atoms in the optical lattice.
From the spectroscopy, the laser linewidth was deduced to be less than 320~Hz, revealing the usefulness of the laser system.

\begin{acknowledgments}
We are grateful to A. Yamaguchi, N. Nemits, D. Akamatsu, and M. Yasuda for the detailed information regarding their clock lasers. 
We would like to acknowledge K. Nakagawa for the mechanical design of the IFDL. 
This work was supported by the Grant-in-Aid for Scientic Research of MEXT/JSPSKAKENHI (No. 25220711, No. 17H06138, No. 18H05405, and No. 18H05228), the Impulsing Paradigm Change through Disruptive Technologies (ImPACT) program, JST CREST (No. JPMJCR1673), and MEXT Q-LEAP. 
\end{acknowledgments}

\nocite{*}
\bibliography{reference}

\end{document}